\documentclass[prl,twocolumn,amsmath,a4paper]{revtex4}

\usepackage{graphicx,amssymb}

\begin{document}
\title{
Universal metallic and insulating properties  \\
of one dimensional Anderson Localization :\\
 a numerical Landauer study}

\author{Guillaume Paulin}
\affiliation{CNRS  UMR 5672 - Laboratoire de Physique de l'Ecole Normale
Sup{\'e}rieure de Lyon, \\
46, All{\'e}e d'Italie, 69007 Lyon, France}

\author{David Carpentier}
\affiliation{CNRS UMR 5672 - Laboratoire de Physique de l'Ecole Normale
Sup{\'e}rieure de Lyon, \\
46, All{\'e}e d'Italie, 69007 Lyon, France}

\date{\today}

\begin{abstract}
We present results on the Anderson localization in a quasi one-dimensional metallic wire in the presence of 
 magnetic impurities. We focus within the same numerical analysis on both the universal localized and metallic regimes, 
 and we study the evolution of these 
 universal properties as the strength of the magnetic disorder is varied. For this purpose, we use 
 a numerical Landauer approach, and derive the scattering matrix of the wire from electron's Green's function 
 obtained from a recursive algorithm.
\end{abstract}

\maketitle

 Interplay between disorder and quantum interferences leads to one of the most remarkable phenomenon in 
 condensed matter : the Anderson localization of waves. The possibility to probe directly the properties of 
 this localization with cold atoms\cite{billy:2008,roati:2008} have greatly renewed the interest on this fascinating physics.  
  In this paper, motivated by the transport properties of metallic spin glass wires 
 \cite{deVegvar:1991,Jaroszynski:1998,Neuttiens:2000}, we study the electronic localization in the presence 
 of both a usual scalar random potential and frozen random magnetic moments.  
 
  One dimensional disordered electronic systems are always localized. Following the scaling theory
  \cite{Abrahams:1979} this implies that by increasing the length $L_{x}$ of the wire for a fixed 
  amplitude of disorder, its typical conductance ultimately reaches  vanishingly small values. The 
  localization length $\xi$ separates metallic regime for small length $L_{x}\ll \xi $ from the 
  asymptotic insulating regime. In the present paper, we focus on several universal properties of 
  both metallic and insulating regime of these wires in the simultaneous presence of two kind of 
  disorder, originating for example from frozen magnetic impurities in a metal. 
    The first type corresponds to scalar potentials induced by the impurities, for which the system 
 has time reversal symmetry (TRS) and spin rotation degeneracy. In this class the Hamiltonian 
 belongs to the so-called  Gaussian Orthogonal Ensemble (GOE) of the Random Matrix Theory 
 classification \cite{Evers:2008} (RMT). 
 If impurities do have a spin, the TRS is broken as well as spin rotation invariance. The Hamiltonian
  is then a unitary matrix, which corresponds in RMT to the Gaussian Unitary Ensemble (GUE) with 
  the breaking of Kramers degeneracy \cite{Mirlin:2000}. However, for the experimentally relevant 
  case of a magnetic potential  weaker than the scalar potential, the system is neither described by the 
  GUE class, nor by the GOE class, but extrapolates in between. 
  
   In this paper, we study numerically the scaling of transport properties of wires with various 
 relative strength of these two disorders. The phase coherence length $L_{\phi}$ which phenomenologically 
 accounts for inelastic scattering\cite{Akkermans:2007} is assumed larger than the wire's length $L_x$. 
 We describe the disordered wire using a tight-binding Anderson lattice model with two kinds of 
 disorder potentials : 
\begin{multline}
\mathcal{H} = t \sum_{<i,j>,s}c_{j,s}^{\dagger}c_{i,s} + \sum_{i,s}v_ic_{i,s}^{\dagger}c_{i,s}  \\
 +  J\sum_{i,s,s'}{\vec{S}}_i .  {\vec{\sigma}}_{s,s'}c_{i,s}^{\dagger}c_{i,s'},
\end{multline}
  The first scalar disorder potential $V=\{v_i\}_i$ is  diagonal in electron-spin space.  
  $v_i$ is a random number  uniformly distributed in the interval $[-W/2,W/2]$. 
  $s,s'$ label the $SU(2)$ spin of electrons and the $\vec{S}_i$ correspond to the frozen classical 
  spin of impurities,  with random orientations uncorrelated from an impurity site to another. 
Varying the  amplitude of magnetic disorder $J$ allows to extrapolate from GOE to GUE. 
For a given realization of disorder, the Landauer conductance of this model on a 2D lattice of size 
(in units of lattice spacing)
$L_{x}\times L_{y}$    is evaluated numerically by a recursive Green's function technique. 
We stay near the band center, avoiding the presence of fluctuating states studied in \cite{Deych:2003}. 
 Universal properties are identified by varying the transverse length $L_y$ from $10$ to $80$, with
the aspect ratio $L_x/L_y$ taken from $1$ to $6000$. Typical number of disorder averages is 5000, 
but for $L_y=10$ we sampled the conductance distribution for $50 000$ different configurations 
of disorder.
 
\paragraph{Localization length.} 

\begin{figure}[ht]
\centerline{\includegraphics[width=9cm,height=5.5cm]{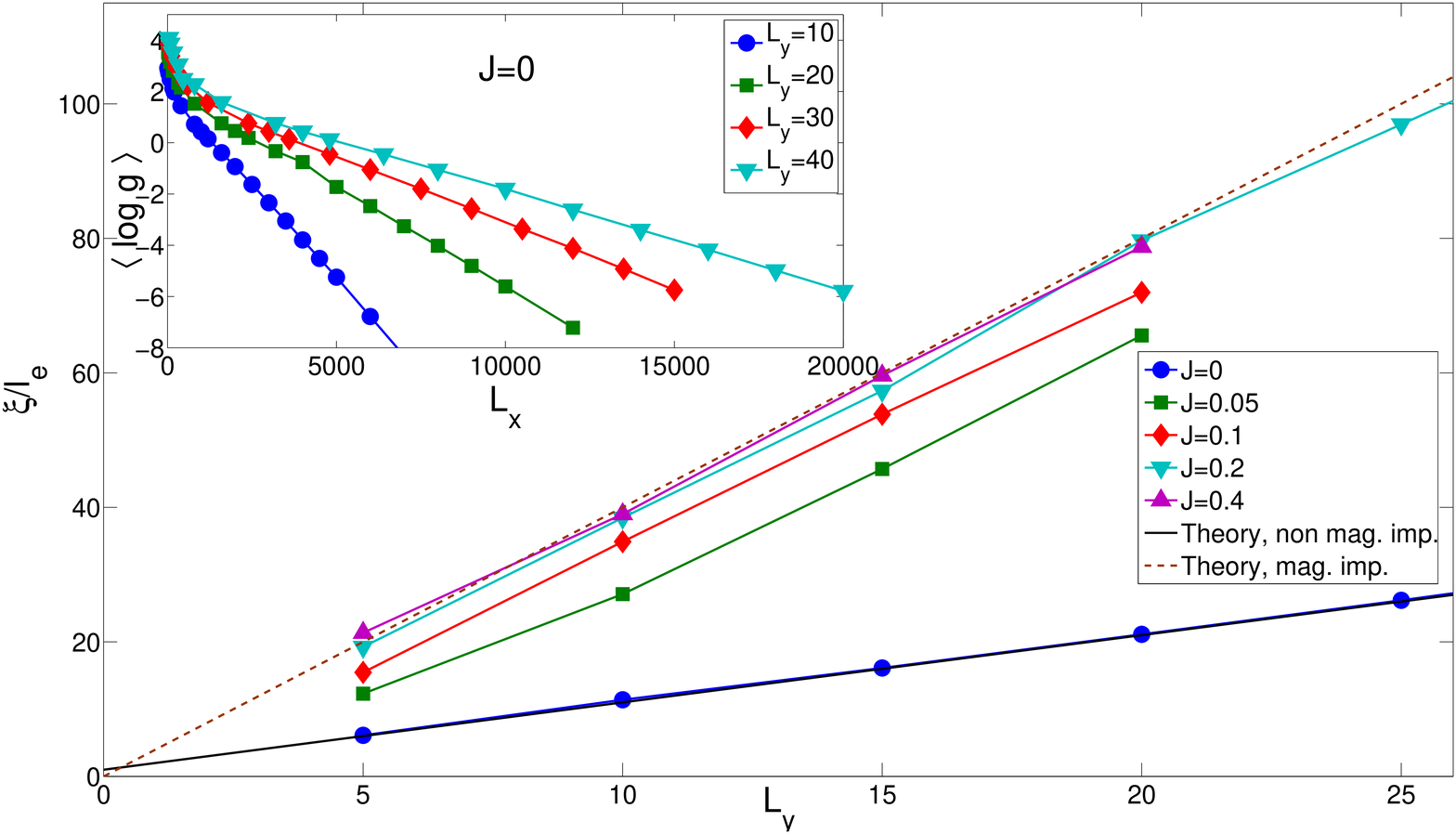}}
\caption{\label{fig:univ_class}
Evolution of localization length as a function of transverse length. $l_e$ is the mean free path of the diffusive sample. Different behavior of the localization length if $J=0$ or $J \neq 0$.
Inset : Scaling of the typical conductance $\langle \log g \rangle = -\frac{2L_x}{\xi}$.
}
\end{figure}

The  localization length $\xi$ is extracted from the scaling of the typical conductance in the localized regime according to\cite{Beenakker:1997,Slevin:2001} :
\begin{equation}
\exp\langle\log g\rangle = \exp\left(-2L_x/\xi\right),
\end{equation}
where $g=G/(e^2/h)$ is the dimensionless conductance and $\langle\cdot\rangle$ represents the 
average over scalar disorder $V$.
 Numerical plot of the logarithm of this equation is given in the inset of figure \ref{fig:univ_class}. The linear behavior of  $\langle \log g \rangle$ is highlighted in the insulating regime and fitted to provide 
the localization length for each value of the transverse length $L_{y}$. Note that we checked that 
 the very slow convergence of the Lyapunov exponent 
$\gamma(L_x) = \frac{1}{2L_x}\log\left(1 + \frac{1}{g(L_x)}\right)$ 
towards $\xi^{-1}(L_{y})$ provided comparable results. 
The evolution of $\xi$ with the transverse length is expected to follow
\cite{Beenakker:1997}:
\begin{equation}
\xi = (\beta L_y + 2 - \beta)l_e, 
\label{equ:xi_gen}
\end{equation}
 with $l_{e}$ the mean free path and $\beta=1$ corresponds to the orthogonal universality class GOE
 while $\beta=2$ for GUE. Note that this change in $\beta$ is accompanied by an artificial 
 doubling of the number of transverse modes $N_y\equiv L_y\to 2N_y$ due to the breaking of Kramers degeneracy \cite{Beenakker:1997}. 
 Comparison of numerical localization lengths for different $J$ with (\ref{equ:xi_gen}) is 
 shown in fig~\ref{fig:univ_class}. 
 Excellent agreement is found for $J=0$ (GOE class, $\beta=1$). In the case $J \neq 0$ we observe 
 a crossover between GOE and GUE for intermediate values of magnetic disorder, while a good agreement 
 with the GUE class is reached for $J\geq 0.2$. 
 From these results, we already notice that the localization regime is reached for much longer 
 wires in the GUE case than for GOE. As shown below, this allows for an easier numerical investigation 
 of the universal metallic regime in the GUE case : magnetic impurities help in finding the universal 
 conductance fluctuations !

\begin{figure}[ht!]
\centerline{\includegraphics[width=9cm,height=5cm]{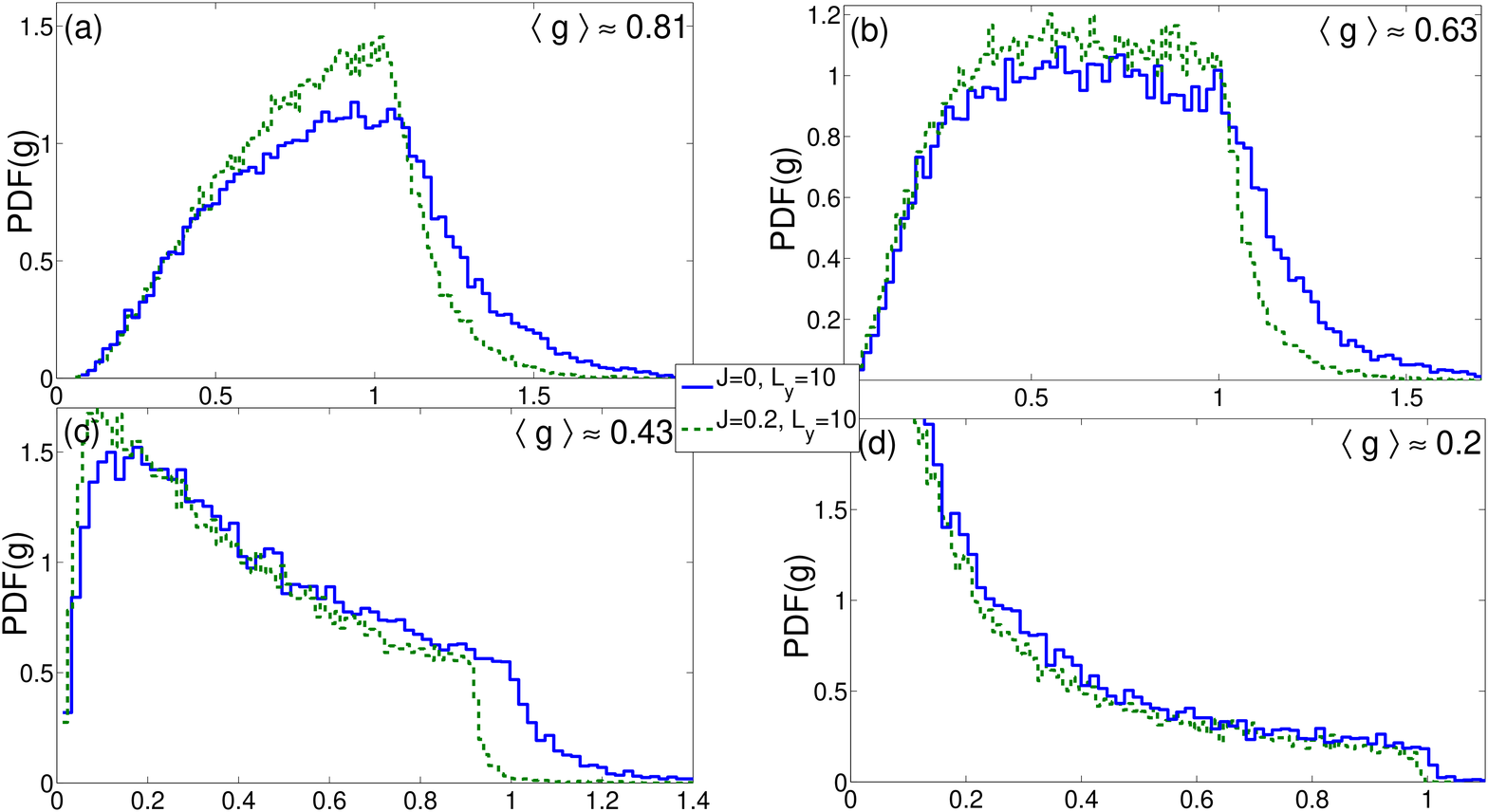}}
\centerline{\includegraphics[width=9cm,height=5cm]{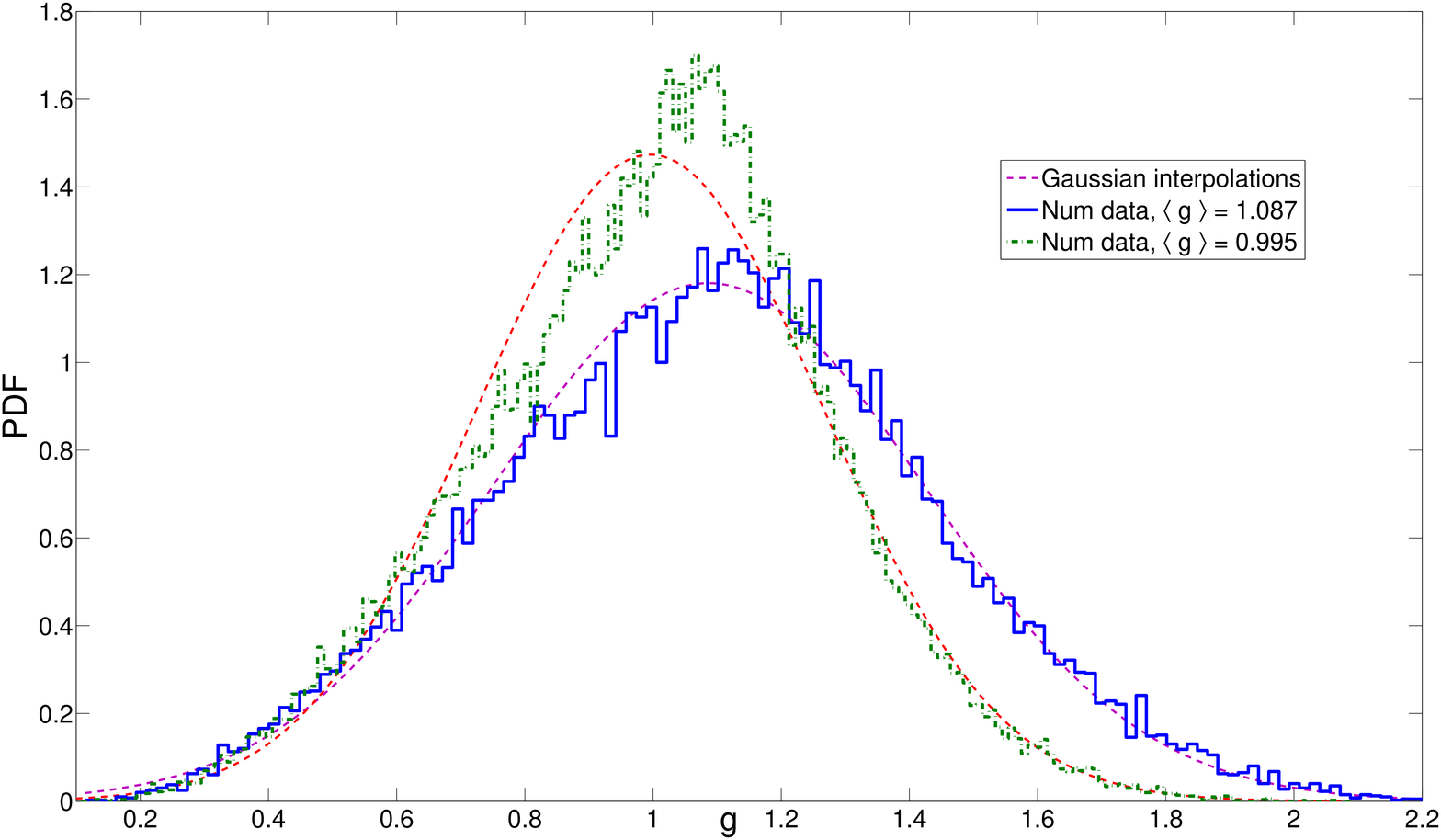}}
\caption{\label{fig:non_analytic}TOP:  Comparison of Probability density functions (PDF) of conductance for $J=0$ (plain curves) and $J=0.2$ (dashed curves). Plots are performed for different values of average conductance. (a):  $\langle g \rangle(J=0)=0.84$ and $\langle g \rangle(J=0.2)=0.79$. (b):  $\langle g \rangle(J=0)=0.67$ and $\langle g \rangle(J=0.2)=0.62$. (c):  $\langle g \rangle(J=0)=0.45$ and $\langle g \rangle(J=0.2)=0.42$. (d):  $\langle g \rangle(J=0)=0.21$ and $\langle g \rangle(J=0.2)=0.18$. BOTTOM:  PDF of conductance for $\langle g \rangle<1 (J=0.2)$ and $\langle g \rangle>1 (J=0)$ and Gaussian interpolations. $L_{y}=10$. 
}
\end{figure}

\begin{figure}[!t]
\includegraphics[width=9cm,height=5.5cm]{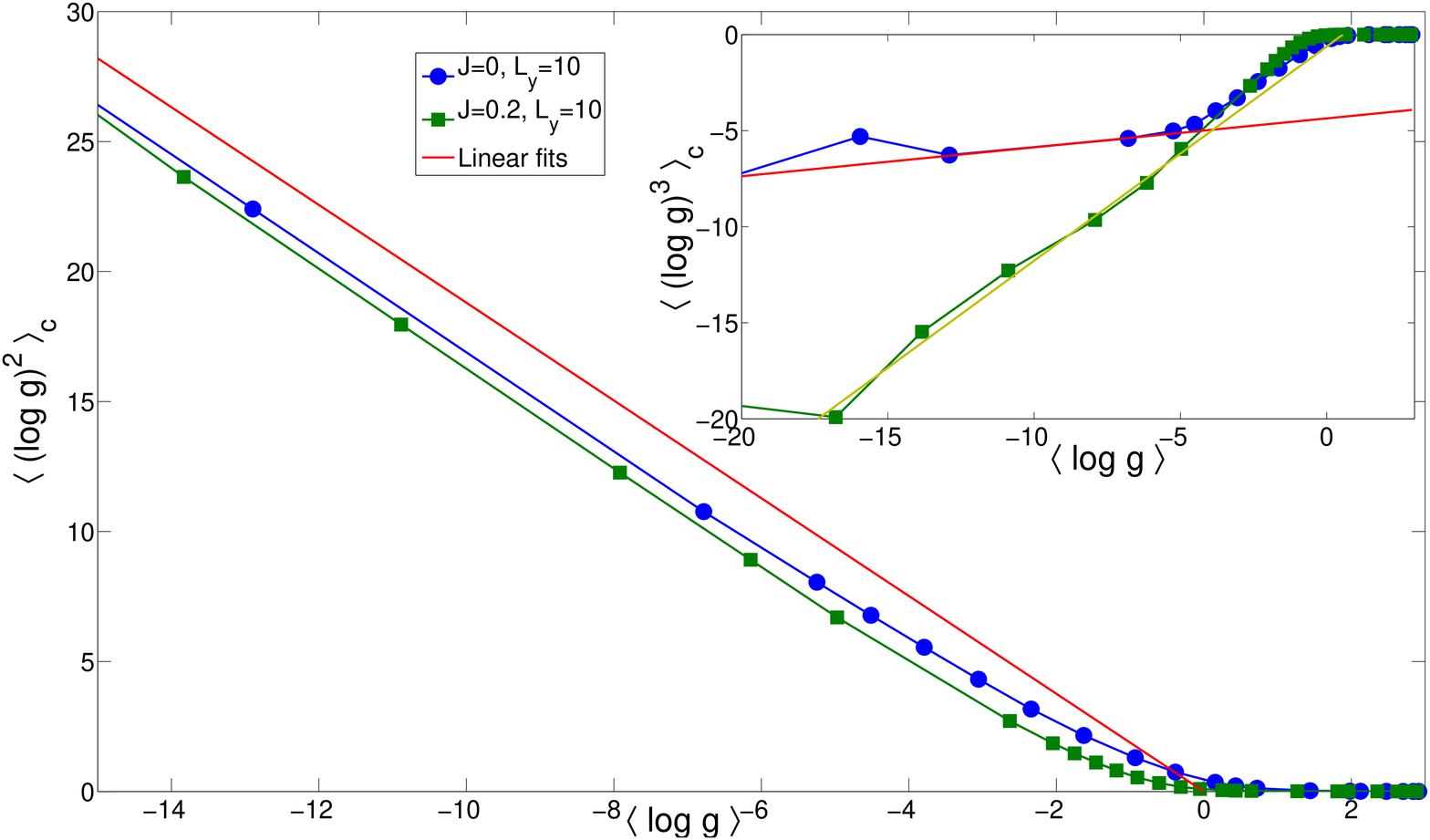}
\caption{\label{fig:logGquad_logGcube}
Plot of the variance of $\log g$ as a function of the mean for the orthogonal and unitary case. In inset is shown the third cumulant of $\log g$ as a function of the mean.
}
\end{figure}

\paragraph{Insulating regime.}
 In the insulating regime $L_{x}\geq \xi $, we expect a Log-normal conductance statistical distribution
 \cite{Beenakker:1997}. 
  However in the region $g=1$ and for $\langle g \rangle\lesssim 1$, 
  we find a non-analytical behavior of $P(g)$ in agreement with 
  \cite{Muttalib:1999, Markos:2002,Froufe:2002, Muttalib:2003} as shown for instance in 
fig.~\ref{fig:non_analytic}. 
In this figure we plot the distribution $P(g)$ for similar values of $\langle g \rangle$ 
for GOE ($J=0$) and GUE ($J=0.2$). 
 The shapes of these distributions are highly similar if $\langle g \rangle \ll 1$, 
 showing that both distributions tend to become Log-normal with the same cumulants. 
 In the intermediate regime, shapes are symmetry dependent. Moreover the non-analyticity appears 
 for different values of conductance (close to $1$) and the rate of the exponential decay 
 \cite{Muttalib:1999} in the metallic regime seems to differ from one ensemble to the other 
 (see for instance curves (a) or (b)). Finally, the bottom curve of 
 figure \ref{fig:non_analytic} represents the distribution of conductance for just above and below the 
 threshold $\langle g \rangle =1$. Plain lines represent gaussian interpolations with a 
 mean and a variance given by the first and the second cumulant of each numerical conductance distribution. For $\langle g \rangle>1$, the gaussian interpolation approximates very well the full 
 distribution. On the other hand, as soon as $\langle g \rangle<1$, the gaussian law only approximates the tail $g\geq 1$ 
 of the distribution of conductance . 
 This behavior is in agreement with the sudden appearance of the non-analyticity for distributions 
 with average conductance inferior to $1$ \cite{Muttalib:2003}. 
 This conductance distribution converges to the Log-normal only deep in the insulating regime, the convergence being very slow (much slower than in the metallic regime). This qualitative result is 
 confirmed by the study of moments :  in the insulating regime the second cumulant is expected 
 to follow\cite{Beenakker:1991}:
\begin{equation}
\langle\left(\log g - \langle\log g\rangle\right)^2\rangle = \langle (\log g)^2\rangle_c = -2\langle\log g\rangle,
\label{eq:pasdaccord}
\end{equation}
Our numerical results are in agreement with this scaling (figure \ref{fig:logGquad_logGcube}) with 
however very slow convergence towards this law :  corrections are measurable even if 
the system is deeply in the localized state. More precisely, we find (see fig.~\ref{fig:logGquad_logGcube}) that for the deep insulating regime 
$\langle (\log g)^2\rangle_c = -1.88 \langle\log g \rangle $ slope $-1.88$, with a slight discrepancy with 
(\ref{eq:pasdaccord}).  
Finally in the inset of figure \ref{fig:logGquad_logGcube}, we show the third cumulant of $\log g$ as a function of the first one. The linear behavior is in agreement with the single parameter scaling. 
We find that 
contrary to the second cumulant the coefficient of proportionality between the skewness and the average depends on the symmetry of disorder, which is not expected.

\begin{figure}[!t]
\includegraphics[width=9cm,height=5.5cm]{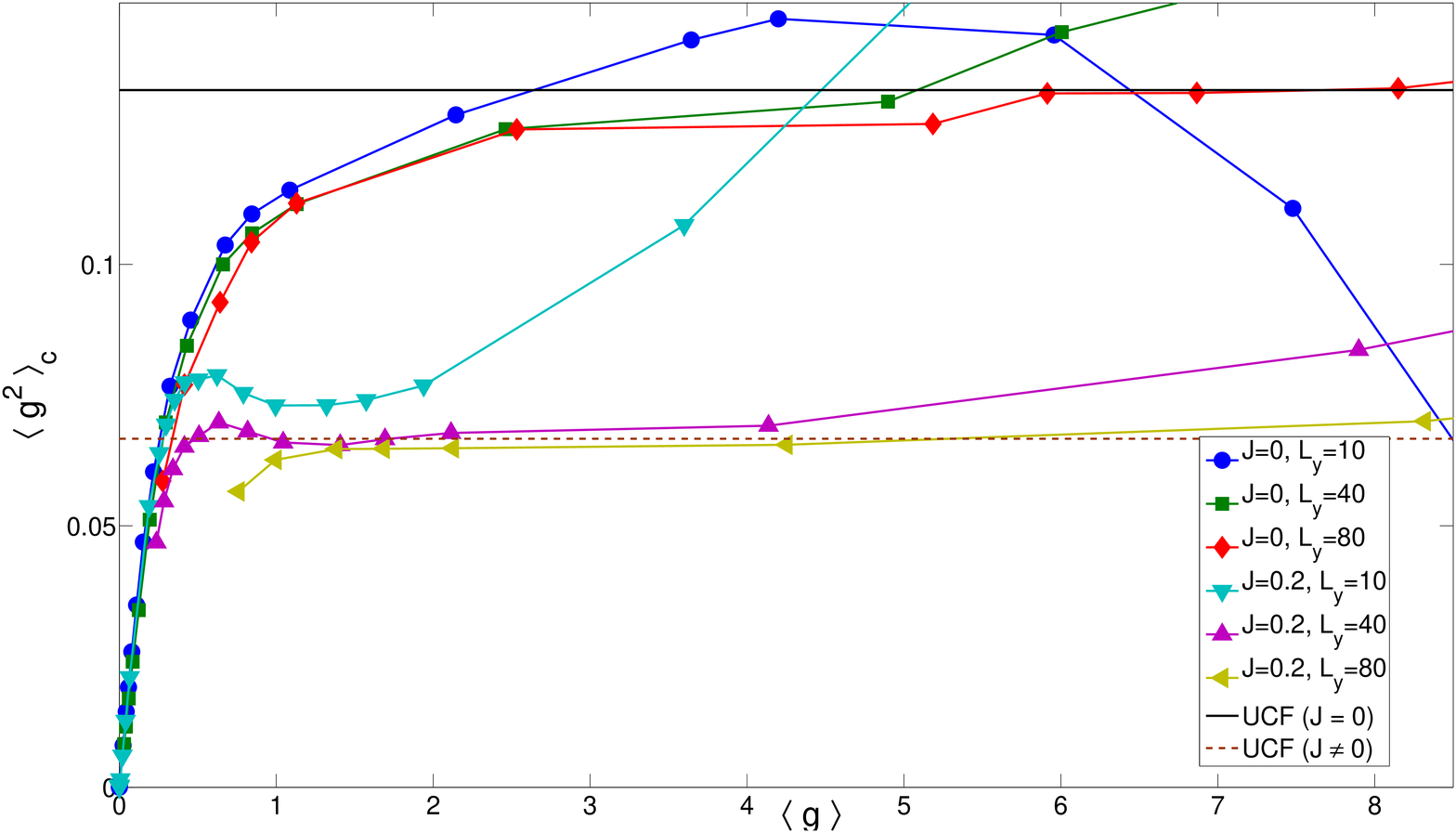}
\includegraphics[width=9cm]{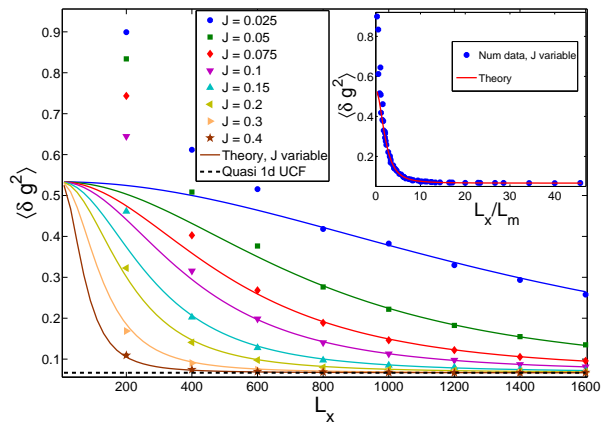}
\caption{\label{fig:Gquad}
TOP : second cumulant of $g$ as a function of first cumulant of $g$ showing the universal behavior in the metallic regime. BOTTOM : variance of $g$ as a function of longitudinal size. UCF are shown. Different curves correspond to different values of magnetic disorder $J$. In inset variance of $g$ is plotted as a function of $L_x/L_m$. Transverse length $L_y=40$.
}
\end{figure}

\paragraph{Metallic regime.}
 We now focus on the universal metallic regime described by weak localization. By definition weak localization  corresponds to metallic diffusion, expected for lengths of wire $l_{e}\ll L_{x}\ll \xi$. 
 For this regime to be reached, we thus need to increase the number of transverse modes $L_{y}$ 
 and thus $\xi$  for all other parameters fixed (see (\ref{equ:xi_gen})). Moreover, for a fixed 
 geometry, this regime will be easier to reach in the GUE class than in the GOE.  
 Fig.~\ref{fig:non_analytic} shows that the conductance distribution is in good approximation Gaussian with a
  variance described \cite{Akkermans:2007} by 
\begin{multline}\label{equ:theo_ucf}
\langle\delta g^2\rangle = \langle g^2 \rangle_c =  \frac{1}{4}F\left(0\right) + \frac{3}{4}F\left(x\sqrt{\frac{4}{3}}\right)  \\
 + \frac{1}{4}F\left(x\sqrt{2}\right) + \frac{1}{4}F\left(x\sqrt{\frac{2}{3}}\right),
\end{multline}
where $x = L_x/L_m$ and the scaling function $F(x)$ depends only on dimension\cite{Akkermans:2007,Paulin:2009b}.
In the bottom plot of figure \ref{fig:Gquad}, these conductance fluctuations are plotted as a function of longitudinal length $L_{x}$ 
for different values of $J$.  A single parameter fit  by (\ref{equ:theo_ucf}) provides the determination of the magnetic dephasing length $L_m$ as a function of the magnetic disorder $J$.  The determination of $L_{m}$ allows for 
very fine comparison with weak localization theory in this regime, allowing for example the study of conductance 
correlation between different disorder configurations (see  \cite{Carpentier:2008,Paulin:2009b}). 
The inset of Fig.~\ref{fig:Gquad} shows the scaling form of these 
fluctuations (as a function of $L_{x}/L_{m}(J)$) in excellent agreement with the theory (\ref{equ:theo_ucf}). 
 Moreover,  for long wires (and large values of $J$) conductance fluctuations are no longer $L_x$ dependent and equal to $1/15$. This is the so-called Universal Conductance Fluctuations (UCF) regime which is precisely identified numerically in the present work.
The top plot of figure \ref{fig:Gquad} confirms analytical results from \cite{Froufe:2002}
 both qualitatively in the shape of the curves and quantitatively in the values of fluctuations in both universality classes. 
  In our study, values of UCF are reached with a maximal error of $1 \%$ for GOE and $3 \%$ for GUE with respect to the analytical value of the UCF in the regime independent  of $\langle g \rangle$ ({\it i.e} with much higher precision 
 than {\it e.g} \cite{Cieplak:1995} and \cite{Markos:2002})  . 
 The other information provided by this curve is the condition for having a universal behavior, \textit{i.e} 
  independence  on the geometry or on the disorder of the sample. 
  As the localized regime is harder to reach for GUE, the universal metallic regime is easier to obtain, 
  the plateau of UCF is widen. One needs then larger transverse lengths (recall that $\xi$ increases 
  with $L_y$) to get the UCF regime for GOE, this is exactly what we see on figure \ref{fig:Gquad} where the UCF plateau is reached only for $L_y=80$ for GOE whereas it is reach for $L_y=80$ and $L_y=40$ for GUE.\\
Finally we consider the third cumulant of the distribution of conductance.  
 According to the analytical study of \cite{Froufe:2002}, this cumulant decays to zero in a universal way
  as $\langle g\rangle$ increases. 
  Here we find a dependance of this decrease on the symmetry class :  for GOE $\langle g^3 \rangle_c$ goes to zero in a monotonous way whereas it decreases, changes its sign and then goes to zero in GUE case.  For $\langle g \rangle>4$ numerical errors are dominant, then this part of the curve is irrelevant.
\begin{figure}[!t]
\includegraphics[width=9cm,height=5.5cm]{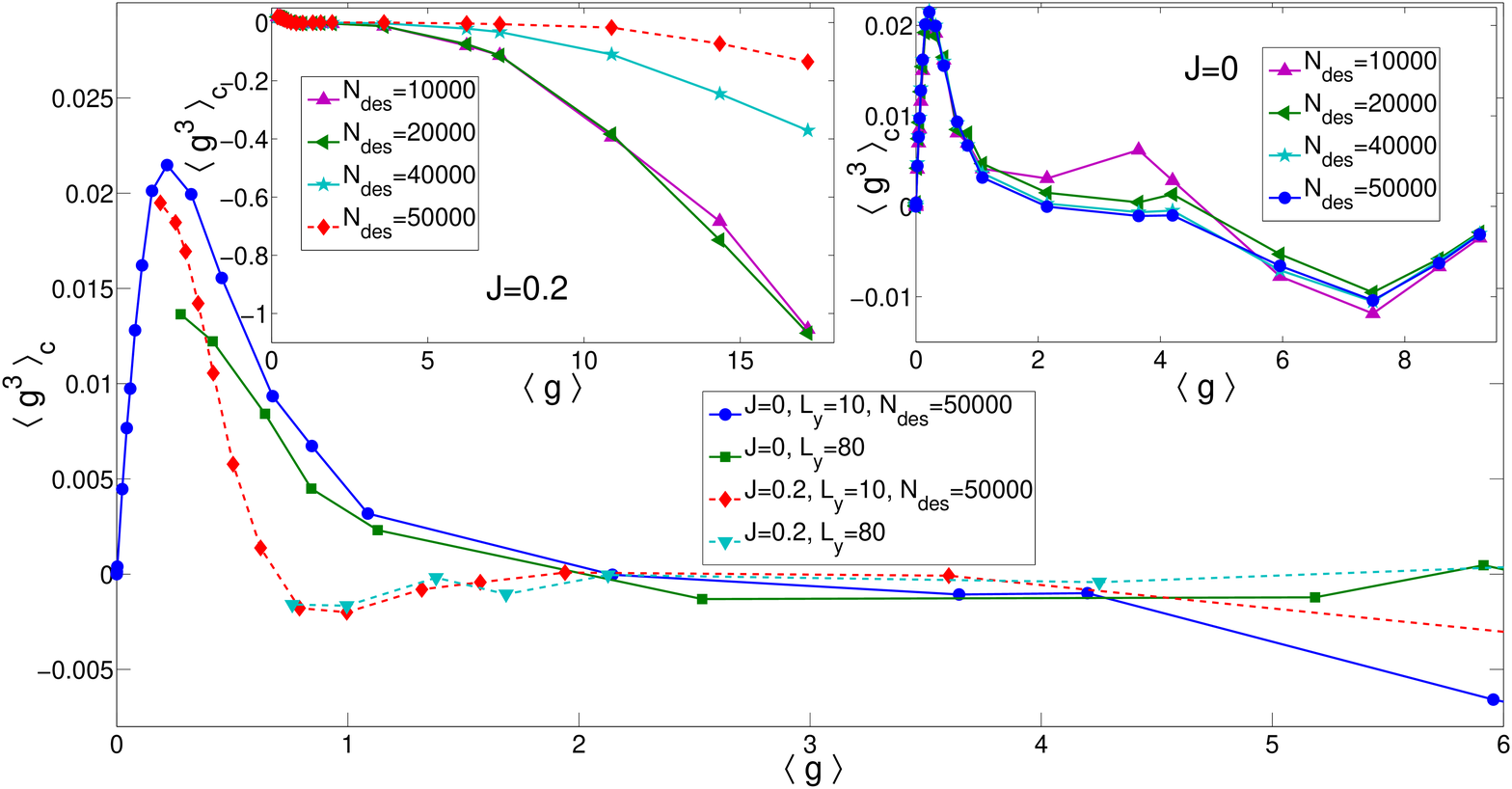}
\caption{\label{fig:Gcube}
Plot of $\langle g^3 \rangle_c$ as a function of $\langle g \rangle$ in the metallic regime. In insets are shown convergence curves for $L_y=10$ and $J=0$ or $J=0.2$.}
\end{figure}
 Note that this fast vanishing of the third cumulant confirms the faster convergence of the whole 
 distribution towards the gaussian, compared to what happens in the insulating regime. 
  Based on our numerical results, we cannot confirm nor refute the expected 
law $\langle g^3 \rangle_c \propto 1/\langle g \rangle^n$, with $n=2$ in GOE and $n=3$ in GUE \cite{Macedo:1994,vanRossum:1997}.

To conclude we have conducted extensive numerical studies of electronic transport in the presence of 
random frozen magnetic moments. Comparing and extending previous analytical and numerical studies, we 
have identified the insulating and metallic regimes described by the universality classes GOE and GUE. We have 
paid special attention to the dependance on this symmetry of cumulants of the distribution of conductance in 
both metallic and insulating universal regimes. In particular, we have identified with high accuracy the 
domain of universal conductance fluctuations, and determined its extension in the present model. 

We thank X. Waintal for useful discussions.
This work was supported by the  ANR grants QuSpins and Mesoglass. All numerical calculations 
were performed on the computing facilities of the ENS-Lyon calculation center (PSMN).  

Note : during the very last step of completion of this paper, we became aware of a preprint 
by Z. Qiao {\it et al.}  \cite{Qiao:2009} which performed a similar numerical Landauer study of 
1D transport for various universality classes, and focused mostly on the metallic regime. While both 
studies agree on the finding of UCF (although we have higher accuracy for $\beta =1$), 
we did not find signs of a second universal plateau for $\langle (\delta g)^2 \rangle$ in our 
study.


\begin{thebibliography}{23}
\expandafter\ifx\csname natexlab\endcsname\relax\def\natexlab#1{#1}\fi
\expandafter\ifx\csname bibnamefont\endcsname\relax
  \def\bibnamefont#1{#1}\fi
\expandafter\ifx\csname bibfnamefont\endcsname\relax
  \def\bibfnamefont#1{#1}\fi
\expandafter\ifx\csname citenamefont\endcsname\relax
  \def\citenamefont#1{#1}\fi
\expandafter\ifx\csname url\endcsname\relax
  \def\url#1{\texttt{#1}}\fi
\expandafter\ifx\csname urlprefix\endcsname\relax\def\urlprefix{URL }\fi
\providecommand{\bibinfo}[2]{#2}
\providecommand{\eprint}[2][]{\url{#2}}

\bibitem[{\citenamefont{Billy et~al.}(2008)\citenamefont{Billy, Josse, Zuo,
  Bernard, Hambrecht, Lugan, Cl{\'e}ment, Sanchez-Palencia, Bouyer, and
  Aspect}}]{billy:2008}
\bibinfo{author}{\bibfnamefont{J.}~\bibnamefont{Billy}},
  \bibinfo{author}{\bibfnamefont{V.}~\bibnamefont{Josse}},
  \bibinfo{author}{\bibfnamefont{Z.}~\bibnamefont{Zuo}},
  \bibinfo{author}{\bibfnamefont{A.}~\bibnamefont{Bernard}},
  \bibinfo{author}{\bibfnamefont{B.}~\bibnamefont{Hambrecht}},
  \bibinfo{author}{\bibfnamefont{P.}~\bibnamefont{Lugan}},
  \bibinfo{author}{\bibfnamefont{D.}~\bibnamefont{Cl{\'e}ment}},
  \bibinfo{author}{\bibfnamefont{L.}~\bibnamefont{Sanchez-Palencia}},
  \bibinfo{author}{\bibfnamefont{P.}~\bibnamefont{Bouyer}}, \bibnamefont{and}
  \bibinfo{author}{\bibfnamefont{A.}~\bibnamefont{Aspect}},
  \bibinfo{journal}{Nature} \textbf{\bibinfo{volume}{453}},
  \bibinfo{pages}{891} (\bibinfo{year}{2008}).

\bibitem[{\citenamefont{Roati et~al.}(2008)\citenamefont{Roati, D'Errico,
  Fallani, Fattori, Fort, Zaccanti, Modugno, Modugno, and
  Inguscio}}]{roati:2008}
\bibinfo{author}{\bibfnamefont{G.}~\bibnamefont{Roati}},
  \bibinfo{author}{\bibfnamefont{C.}~\bibnamefont{D'Errico}},
  \bibinfo{author}{\bibfnamefont{L.}~\bibnamefont{Fallani}},
  \bibinfo{author}{\bibfnamefont{M.}~\bibnamefont{Fattori}},
  \bibinfo{author}{\bibfnamefont{C.}~\bibnamefont{Fort}},
  \bibinfo{author}{\bibfnamefont{M.}~\bibnamefont{Zaccanti}},
  \bibinfo{author}{\bibfnamefont{G.}~\bibnamefont{Modugno}},
  \bibinfo{author}{\bibfnamefont{M.}~\bibnamefont{Modugno}}, \bibnamefont{and}
  \bibinfo{author}{\bibfnamefont{M.}~\bibnamefont{Inguscio}},
  \bibinfo{journal}{Nature} \textbf{\bibinfo{volume}{453}},
  \bibinfo{pages}{895} (\bibinfo{year}{2008}).

\bibitem[{\citenamefont{de~Vegvar et~al.}(1991)\citenamefont{de~Vegvar,
  L{\'e}vy, and Fulton}}]{deVegvar:1991}
\bibinfo{author}{\bibfnamefont{P.}~\bibnamefont{de~Vegvar}},
  \bibinfo{author}{\bibfnamefont{L.}~\bibnamefont{L{\'e}vy}}, \bibnamefont{and}
  \bibinfo{author}{\bibfnamefont{T.}~\bibnamefont{Fulton}},
  \bibinfo{journal}{Phys. Rev. Lett.} \textbf{\bibinfo{volume}{66}},
  \bibinfo{pages}{2380} (\bibinfo{year}{1991}).

\bibitem[{\citenamefont{Jaroszynski et~al.}(1998)\citenamefont{Jaroszynski,
  Wrobel, Karczewski, Wojtowicz, and Dietl}}]{Jaroszynski:1998}
\bibinfo{author}{\bibfnamefont{J.}~\bibnamefont{Jaroszynski}},
  \bibinfo{author}{\bibfnamefont{J.}~\bibnamefont{Wrobel}},
  \bibinfo{author}{\bibfnamefont{G.}~\bibnamefont{Karczewski}},
  \bibinfo{author}{\bibfnamefont{T.}~\bibnamefont{Wojtowicz}},
  \bibnamefont{and} \bibinfo{author}{\bibfnamefont{T.}~\bibnamefont{Dietl}},
  \bibinfo{journal}{Phys. Rev. Lett.} \textbf{\bibinfo{volume}{80}},
  \bibinfo{pages}{5635} (\bibinfo{year}{1998}).

\bibitem[{\citenamefont{Neuttiens et~al.}(2000)\citenamefont{Neuttiens, Strunk,
  Haesendonck, and Bruynseraede}}]{Neuttiens:2000}
\bibinfo{author}{\bibfnamefont{G.}~\bibnamefont{Neuttiens}},
  \bibinfo{author}{\bibfnamefont{C.}~\bibnamefont{Strunk}},
  \bibinfo{author}{\bibfnamefont{C.~V.} \bibnamefont{Haesendonck}},
  \bibnamefont{and}
  \bibinfo{author}{\bibfnamefont{Y.}~\bibnamefont{Bruynseraede}},
  \bibinfo{journal}{Phys. Rev. B} \textbf{\bibinfo{volume}{62}},
  \bibinfo{pages}{3905} (\bibinfo{year}{2000}).

\bibitem[{\citenamefont{Abrahams et~al.}(1979)\citenamefont{Abrahams, Anderson,
  Licciardello, and Ramakrishnan}}]{Abrahams:1979}
\bibinfo{author}{\bibfnamefont{E.}~\bibnamefont{Abrahams}},
  \bibinfo{author}{\bibfnamefont{P.}~\bibnamefont{Anderson}},
  \bibinfo{author}{\bibfnamefont{D.}~\bibnamefont{Licciardello}},
  \bibnamefont{and}
  \bibinfo{author}{\bibfnamefont{T.}~\bibnamefont{Ramakrishnan}},
  \bibinfo{journal}{Phys. Rev. Lett.} \textbf{\bibinfo{volume}{42}},
  \bibinfo{pages}{673} (\bibinfo{year}{1979}).

\bibitem[{\citenamefont{Evers and Mirlin}(2008)}]{Evers:2008}
\bibinfo{author}{\bibfnamefont{F.}~\bibnamefont{Evers}} \bibnamefont{and}
  \bibinfo{author}{\bibfnamefont{A.~D.} \bibnamefont{Mirlin}},
  \bibinfo{journal}{Rev. Mod. Phys.} \textbf{\bibinfo{volume}{80}},
  \bibinfo{pages}{1355} (\bibinfo{year}{2008}).

\bibitem[{\citenamefont{Mirlin}(2000)}]{Mirlin:2000}
\bibinfo{author}{\bibfnamefont{A.~D.} \bibnamefont{Mirlin}},
  \bibinfo{journal}{Phys. Rep.} \textbf{\bibinfo{volume}{326}},
  \bibinfo{pages}{259} (\bibinfo{year}{2000}).

\bibitem[{\citenamefont{Akkermans and Montambaux}(2007)}]{Akkermans:2007}
\bibinfo{author}{\bibfnamefont{E.}~\bibnamefont{Akkermans}} \bibnamefont{and}
  \bibinfo{author}{\bibfnamefont{G.}~\bibnamefont{Montambaux}},
  \emph{\bibinfo{title}{Mesoscopic Physics of electrons and photons}}
  (\bibinfo{publisher}{Cambridge University Press}, \bibinfo{year}{2007}).

\bibitem[{\citenamefont{Deych et~al.}(2003)\citenamefont{Deych, Erementchouk,
  and Lisyansky}}]{Deych:2003}
\bibinfo{author}{\bibfnamefont{L.~I.} \bibnamefont{Deych}},
  \bibinfo{author}{\bibfnamefont{M.~V.} \bibnamefont{Erementchouk}},
  \bibnamefont{and} \bibinfo{author}{\bibfnamefont{A.~A.}
  \bibnamefont{Lisyansky}}, \bibinfo{journal}{Phys. Rev. Lett.}
  \textbf{\bibinfo{volume}{90}}, \bibinfo{pages}{126601}
  (\bibinfo{year}{2003}).

\bibitem[{\citenamefont{Beenakker}(1997)}]{Beenakker:1997}
\bibinfo{author}{\bibfnamefont{C.~W.~J.} \bibnamefont{Beenakker}},
  \bibinfo{journal}{Rev. Mod. Phys.} \textbf{\bibinfo{volume}{69}},
  \bibinfo{pages}{731} (\bibinfo{year}{1997}).

\bibitem[{\citenamefont{Slevin et~al.}(2001)\citenamefont{Slevin, Markos, and
  Ohtsuki}}]{Slevin:2001}
\bibinfo{author}{\bibfnamefont{K.}~\bibnamefont{Slevin}},
  \bibinfo{author}{\bibfnamefont{P.}~\bibnamefont{Markos}}, \bibnamefont{and}
  \bibinfo{author}{\bibfnamefont{T.}~\bibnamefont{Ohtsuki}},
  \bibinfo{journal}{Phys. Rev. Lett.} \textbf{\bibinfo{volume}{86}},
  \bibinfo{pages}{3594} (\bibinfo{year}{2001}).

\bibitem[{\citenamefont{Muttalib and W{\"o}lfle}(1999)}]{Muttalib:1999}
\bibinfo{author}{\bibfnamefont{K.~A.} \bibnamefont{Muttalib}} \bibnamefont{and}
  \bibinfo{author}{\bibfnamefont{P.}~\bibnamefont{W{\"o}lfle}},
  \bibinfo{journal}{Phys. Rev. Lett.} \textbf{\bibinfo{volume}{83}},
  \bibinfo{pages}{3013} (\bibinfo{year}{1999}).

\bibitem[{\citenamefont{Markos}(2002)}]{Markos:2002}
\bibinfo{author}{\bibfnamefont{P.}~\bibnamefont{Markos}},
  \bibinfo{journal}{Phys. Rev. B} \textbf{\bibinfo{volume}{65}},
  \bibinfo{pages}{104207} (\bibinfo{year}{2002}).

\bibitem[{\citenamefont{Froufe-P{\'e}rez
  et~al.}(2002)\citenamefont{Froufe-P{\'e}rez, Garcia-Mochales, Serena, Mello,
  and Saenz}}]{Froufe:2002}
\bibinfo{author}{\bibfnamefont{L.}~\bibnamefont{Froufe-P{\'e}rez}},
  \bibinfo{author}{\bibfnamefont{P.}~\bibnamefont{Garcia-Mochales}},
  \bibinfo{author}{\bibfnamefont{P.}~\bibnamefont{Serena}},
  \bibinfo{author}{\bibfnamefont{P.}~\bibnamefont{Mello}}, \bibnamefont{and}
  \bibinfo{author}{\bibfnamefont{J.}~\bibnamefont{Saenz}},
  \bibinfo{journal}{Phys. Rev. Lett.} \textbf{\bibinfo{volume}{89}},
  \bibinfo{pages}{246403} (\bibinfo{year}{2002}).

\bibitem[{\citenamefont{Muttalib et~al.}(2003)\citenamefont{Muttalib,
  W{\"o}lfle, Garc{\'\i}a-Mart{\'\i}n, and Gopar}}]{Muttalib:2003}
\bibinfo{author}{\bibfnamefont{K.~A.} \bibnamefont{Muttalib}},
  \bibinfo{author}{\bibfnamefont{P.}~\bibnamefont{W{\"o}lfle}},
  \bibinfo{author}{\bibfnamefont{A.}~\bibnamefont{Garc{\'\i}a-Mart{\'\i}n}},
  \bibnamefont{and} \bibinfo{author}{\bibfnamefont{V.~A.} \bibnamefont{Gopar}},
  \bibinfo{journal}{Europhys. Lett.} \textbf{\bibinfo{volume}{61}},
  \bibinfo{pages}{95} (\bibinfo{year}{2003}).

\bibitem[{\citenamefont{Beenakker and van Houten}(1991)}]{Beenakker:1991}
\bibinfo{author}{\bibfnamefont{C.~W.~J.} \bibnamefont{Beenakker}}
  \bibnamefont{and} \bibinfo{author}{\bibfnamefont{H.}~\bibnamefont{van
  Houten}}, \bibinfo{journal}{Solid State Physics}
  \textbf{\bibinfo{volume}{44}}, \bibinfo{pages}{1} (\bibinfo{year}{1991}).

\bibitem[{\citenamefont{Paulin and Carpentier}(2009)}]{Paulin:2009b}
\bibinfo{author}{\bibfnamefont{G.}~\bibnamefont{Paulin}} \bibnamefont{and}
  \bibinfo{author}{\bibfnamefont{D.}~\bibnamefont{Carpentier}}
  (\bibinfo{year}{2009}), \bibinfo{note}{preprint}.

\bibitem[{\citenamefont{Carpentier and Orignac}(2008)}]{Carpentier:2008}
\bibinfo{author}{\bibfnamefont{D.}~\bibnamefont{Carpentier}} \bibnamefont{and}
  \bibinfo{author}{\bibfnamefont{E.}~\bibnamefont{Orignac}},
  \bibinfo{journal}{Phys. Rev. Lett.} \textbf{\bibinfo{volume}{100}},
  \bibinfo{pages}{057207} (\bibinfo{year}{2008}).

\bibitem[{\citenamefont{Cieplak et~al.}(1995)\citenamefont{Cieplak, Bulka, and
  Dietl}}]{Cieplak:1995}
\bibinfo{author}{\bibfnamefont{M.}~\bibnamefont{Cieplak}},
  \bibinfo{author}{\bibfnamefont{B.}~\bibnamefont{Bulka}}, \bibnamefont{and}
  \bibinfo{author}{\bibfnamefont{T.}~\bibnamefont{Dietl}},
  \bibinfo{journal}{Phys. Rev. B} \textbf{\bibinfo{volume}{51}},
  \bibinfo{pages}{8939} (\bibinfo{year}{1995}).

\bibitem[{\citenamefont{Mac{\^e}do}(1994)}]{Macedo:1994}
\bibinfo{author}{\bibfnamefont{A.~M.~S.} \bibnamefont{Mac{\^e}do}},
  \bibinfo{journal}{Phys. Rev. B} \textbf{\bibinfo{volume}{49}},
  \bibinfo{pages}{1858} (\bibinfo{year}{1994}).

\bibitem[{\citenamefont{van Rossum et~al.}(1997)\citenamefont{van Rossum,
  Lerner, Altshuler, and Nieuwenhuizen}}]{vanRossum:1997}
\bibinfo{author}{\bibfnamefont{M.}~\bibnamefont{van Rossum}},
  \bibinfo{author}{\bibfnamefont{I.~V.} \bibnamefont{Lerner}},
  \bibinfo{author}{\bibfnamefont{B.~L.} \bibnamefont{Altshuler}},
  \bibnamefont{and} \bibinfo{author}{\bibfnamefont{T.~M.}
  \bibnamefont{Nieuwenhuizen}}, \bibinfo{journal}{Phys. Rev. B}
  \textbf{\bibinfo{volume}{55}}, \bibinfo{pages}{4710} (\bibinfo{year}{1997}).

\bibitem[{\citenamefont{Qiao et~al.}()\citenamefont{Qiao, Xing, and
  Wang}}]{Qiao:2009}
\bibinfo{author}{\bibfnamefont{Z.}~\bibnamefont{Qiao}},
  \bibinfo{author}{\bibfnamefont{Y.}~\bibnamefont{Xing}}, \bibnamefont{and}
  \bibinfo{author}{\bibfnamefont{J.}~\bibnamefont{Wang}},
  \bibinfo{note}{arXiv:0910.3475v1}.

\end{thebibliography}

\end{document}